\documentclass[amsmath,amssymb,preprint,goupaddress,prb,aps,showpacs,showkeys]{revtex4}
\usepackage{bm,graphicx,subfigure,enumerate,latexsym}

\begin{document}
\title{A Model for Hybrid Simulations of Molecular Dynamics and CFD}
\author{Shugo Yasuda
\footnote{Electronic mail: yasuda@cheme.kyoto-u.ac.jp}}
\author{Ryoichi Yamamoto
\footnote{Electronic mail: ryoichi@cheme.kyoto-u.ac.jp}}
\affiliation{
Department of Chemical Engineering, 
Kyoto University, Kyoto 615-8510, Japan
and
CREST, Japan Science and Technology Agency, Kawaguchi 332-0012, Japan.
}
\date{\today}

\begin{abstract}
We propose a method for multi-scale hybrid simulations of molecular
 dynamics (MD) and computational fluid dynamics (CFD). 
In the method, usual lattice-mesh based simulations are applied for CFD
 level, but each lattice is associated with a small MD cell which
 generates a ``local stress'' according to a ``local flow field'' given
 from CFD instead of using any constitutive functions at CFD level. 
We carried out the hybrid simulations for some elemental flow problems
 of simple Lennard-Jones liquids and compared the results with those
 obtained by usual CFDs with a Newtonian constitutive relation in order
 to examine the validity of our hybrid simulation method. 
It is demonstrated that our hybrid simulations successfully reproduced
 the correct flow behavior obtained from usual CFDs as far as the mesh
 size $\Delta x$ and the time-step $\Delta t$ of CFD are not too large
 comparing to the system size $l_{\rm MD}$ and the sampling duration
 $t_{\rm MD}$ of MD simulations performed at each time step of CFDs.
Otherwise, simulations are affected by large fluctuations due to poor
 statistical averages taken in the MD part. 
Properties of the fluctuations are analyzed in detail.
\end{abstract}


\maketitle

\section{Introduction}
Hydrodynamics of complex fluids are of particular importance in
various science and engineering fields, such as fluid mechanics,
soft matter science, mechanical engineering, chemical engineering, and
so on.
Because of the complicated couplings between internal degree of freedoms
of complex fluids and their flow behavior, conventional treatments, 
based on usual
assumptions such as non-slip boundary conditions and linear Newtonian
constitutive relations,
are often invalid. 
Striking examples can be seen in systems such as colloidal
dispersions, polymeric liquids, granular matters, and liquid crystals.
Those systems are known to exhibit peculiar flow behaviors, e.g., shear
thinning or thickening, viscoelasticity, jamming, flow induced phase
transition, etc.

Although there exits huge accumulation of experimental and theoretical
studies on the rheology of complex fluids, performing computational
fluid dynamics (CFD) simulations are not yet common for complex fluids 
since reliable constitutive equations are often unknown for those systems.
On the other hand, there exists a different problem also for microscopic
approaches such as molecular dynamics (MD) simulations, 
while constitutive equations are no more necessary in this case.
The characteristic time and length scales of complex fluids easily
become several orders larger than those of microscopic scales. 
Therefore, most hydrodynamic problems of complex fluids are yet out of
reach of microscopic MD simulations. 
To overcome those serious limitations mentioned above, we aim to develop a
new multi-scale method which is for performing hybrid simulations of MD
and CFD valid for complex fluids without any constitutive equations.

Various methods for hybrid simulations of MD and CFD has already 
been proposed by several researchers.
Most of those methods are based on ``domain decomposition'' 
for which MD simulations are applied only around the points
of interest, {\it i.e.}, in the vicinity of defects, boundaries,
interfaces, where details of molecular motions are important, while the
remaining regions are treated only by CFD.
\cite{art:95OT,art:00FWF,art:03DC,art:05DFC,art:04NCER,art:04NCR,art:06NRC,art:07LCNR,art:07R,art:07YST}
Exchange of information between MD and CFD is performed in a coupling 
regions where each system is subjected to some constrains to take the
consistency of the two systems. 
This kind of hybrid method is expected to be useful specially for 
problems including interfaces, such as adhesion, friction, anchoring of
crystal liquids, stick-slip motions, etc.

In order to apply hybrid methods of MD and CFD to hydrodynamics of
complex fluids, a different type of approach is probably needed.
Our strategy for this is straightforward. 
We try to develop a multi-scale hybrid method based on the local equilibrium assumption.
Here CFD is used as a fluid solver, while MD simulations are used only
to generate local properties, such as constitutive relations of the
fluid under consideration, by performing local statistical sampling
in a consistent matter. 
The numerical algorithm is rather simple. 
We perform usual lattice-mesh based CFD simulations at an upper level,
but each mesh-node is associated with a small lower level MD cell which 
passes a ``local stress'' to CFD according to a ``local flow field'' given 
from CFD to MD instead of using any constitutive functions at CFD
level. 
MD simulations thus have to be performed at all node points and at every 
time steps of CFD.

One might think that the simulations would be be much faster 
if we construct tabular database of the constitutive relations by
performing MD simulations in advance under many different simulation parameters 
and refer the table from CFD. 
The ``tabular approach'' works much effective for simple fluids for
which the constitutive relations depend only on a few parameters, such
as the density, temperature, and shear rate. 
In the case of complex fluids, however, the number of parameters to be
considered can be huge depending on the local quantities to be
considered.
In the case of charged systems for example, the local stress depends
also on the local compositions and chemical potentials of ions and
the local electric field, etc. 
Although we used only simple Lennard-Jones liquid in the present study, 
we adopt the local sampling strategy rather than
the tabular contraction strategy to be more general.
The main purpose of the present study is to examine the validity of our
multi-scale hybrid model by performing some simple demonstrations of the
method.
Efficiency and drawback of both strategies will be considered in future for
more specific problems.
An idea similar in spirit to the present method was
also put forward earlier by W. Ren and W. E. \cite{art:05RE}

The hybrid simulation method is described in Sec. II,
and some demonstrative results for one- and two-dimensional flows of
simple Lennard-Jones liquids are shown in Sec. III. 
A special attention is put on the efficiency and the reliability of 
our hybrid method there. 
The simulation results obtained by our multi-scale hybrid method are
compared with those of normal CFDs with a Newtonian constitutive
relation.
The validity of our method is discussed in Sec. IV, and a summary is
given in Sec. V.

\section{Hybrid Model}
Incompressible flows for isotropic materials are described by the following equations,
\begin{equation}\label{eq1}
\frac{\partial v_\alpha}{\partial x_\alpha} = 0,
\end{equation}
\begin{equation}\label{eq2}
\frac{\partial v_\alpha}{\partial t} + v_\beta\frac{\partial v_\alpha}{\partial x_\beta}
=\frac{1}{\rho}\frac{\partial P_{\alpha\beta}}{\partial x_\beta} + g_\alpha,
\end{equation}
where $x_\alpha$ is the Cartesian coordinate system, $t$ the time, $v_\alpha$ the velocity, $\rho$ the density, $P_{\alpha\beta}$ the stress tensor,
and $g_\alpha$ the external force per unit mass. Here and after the subscripts $\alpha$, $\beta$, and $\gamma$ represent the index in Cartesian coordinates, i.e.
\{$\alpha$, $\beta$, $\gamma$\} = \{$x$, $y$, $z$\}, and the summation convention is used.
The stress tensor $P_{\alpha\beta}$ is written in the form,
\begin{equation}\label{eq3}
P_{\alpha\beta} = -p \delta_{\alpha\beta} + T_{\alpha\beta},
\end{equation}
where $p$ is the pressure and $\delta_{\alpha\beta}$ is the Kronecker delta.
Here we assumed that the diagonal component of the stress tensor is isotropic. The off-diagonal stress tensor is symmetric $T_{\alpha\beta}=T_{\beta\alpha}$ and traceless $T_{\alpha\alpha}$=0. \cite{art:45R}
In order to solve the above equations, one needs a constitutive relation for the stress tensor $T_{\alpha\beta}$. In our hybrid method, instead of using any explicit formulas such as the Newtonian constitutive relation, $T_{\alpha\beta}$ is computed directly by MD simulations.

\subsection{CFD Scheme}
\begin{figure}[t]
\begin{center}
\includegraphics[scale=1]{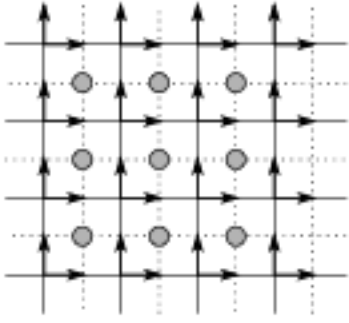}
\end{center}
\caption{Staggered arrangement of vector quantity, the velocity $\bm v$, and scalar quantity, the pressure $p$ and density $\rho$, on a lattice-mesh grid.
}
\label{f0}
\end{figure}

We use a lattice-mesh based finite volume method with a staggered arrangement for vector and scalar quantity.\cite{book:02FP}  See Fig.~\ref{f0}. The control volume for a vector quantity is a unit square surrounded by dashed lines and that for a scalar quantity is a unit square surrounded by solid lines.
Eqs (\ref{eq1}) and (\ref{eq2}) are discretized by integrating the quantities on each control volume. As for numerical time integrations, we use the fourth order Runge-Kutta method, where a single physical time step $\Delta t$ is divided into four sub-steps. More concretely, the time evolution of a quantity $\phi$, which is to be determined by the equation $\partial \phi/\partial t$=$f(t,\phi)$, is written as
\begin{subequations}\label{eq4}
\begin{align}
\phi_{n+\frac{1}{2}}^*&=\phi^n + \frac{\Delta t}{2}f(t_n,\phi^n),
\\
\phi_{n+\frac{1}{2}}^{**}&=\phi^n + \frac{\Delta t}{2}f(t_{n+\frac{1}{2}},\phi^*_{n+\frac{1}{2}}),
\\
\phi_{n+1}^{*}&=\phi^n + \Delta t f(t_{n+\frac{1}{2}},\phi^{**}_{n+\frac{1}{2}}),
\\
\phi^{n+1}&=\phi^n + \frac{\Delta t}{6}\left[
f(t_{n},\phi^{n}) + 2 f(t_{n+\frac{1}{2}},\phi^*_{n+\frac{1}{2}}) +
\right.
\nonumber
\\
&\hspace{2cm} 
\left. 2 f(t_{n+1/2},\phi^{**}_{n+1/2}) + f(t_{n+1},\phi^*_{n+1}) 
\right].
\end{align}
\end{subequations}
Time evolution of the fluid velocity $\bm v$ is computed by the above set of equations.
On the other hand, the pressure $p$ is determined so that the fluid velocity  satisfies the incompressible condition (\ref{eq1}) at each sub-step. 
The procedure at each sub-step is written as
\begin{subequations}\label{eq5}
\begin{align}
&p = \tilde p + \psi,
\\
&\bm v =\tilde {\bm v} -\tau \nabla \psi,
\\
&\triangle \psi = \frac{1}{\tau}\nabla{\tilde {\bm v}},
\end{align}
\end{subequations}
where $\tilde p$ is the pressure obtained at the previous sub-step, $\tilde {\bm v}$ is the velocity obtained by solving equation (\ref{eq4}) at the present sub-step, and $\tau$ is the time increment of the sub-step. The remaining three components of the tensor $T_{\alpha\beta}$ are to be computed directly by MD simulations. The detail of the method is described in the next subsection.
Note that the calculations of $T_{\alpha\beta}$ is carried out at each sub-step of equation (\ref{eq4}). 

\begin{figure*}[htbp]
\begin{center}
\includegraphics[scale=1]{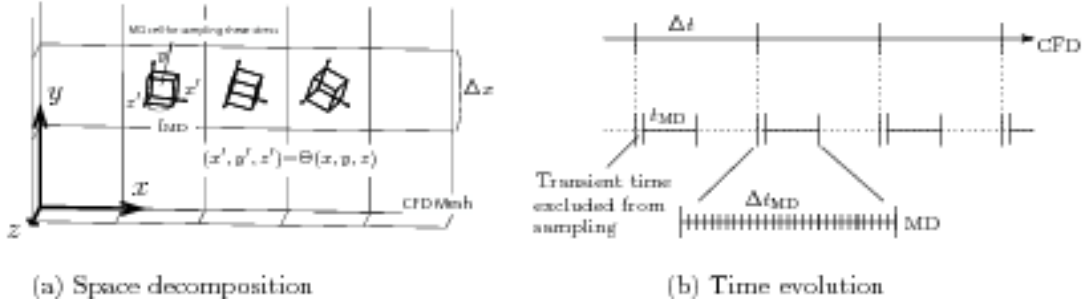}
\end{center}
\caption{Schematic diagram for the hybrid scheme.
(a) CFD simulations are performed in a reference coordinate ($x$,$y$,$z$),
while MD simulations are performed in a rotated coordinate ($x'$,$y'$,$z'$)
so that the diagonal components of $E'_{\alpha\beta}$
become all zero with the procedure described in Sec II B.
The CFD system is discretized into cubic subsystems whose side length is $\Delta x$. Each subsystem is associate with a MD cell, whose side length is
$l_{\rm MD}$, with Lees-Edward periodic boundary condition under shear deformation.
(b) A schematic time evolution of our multi-scale method. CFD simulation
proceeds with a time step of $\Delta t$, MD simulation is carried out for
a lapse of time $t_{\rm MD}$ only to sample local stress $T'_{\alpha\beta}$
at each node point and time step of CFD.
}
\label{f00}
\end{figure*}
\subsection{Computation of Local Stress by MD}
We compute the local stresses by MD simulations according to the local strain rates, rather than the local flow velocities themselves, computed at the CFD level. A schematic diagram of the method is depicted in Fig. \ref{f00}.
At the CFD level, the local strain rate tensor $E_{\alpha\beta}$ is defined as
\begin{equation}\label{eq6}
E_{\alpha\beta}=\frac{1}{2}\left (\frac{\partial v_\alpha}{\partial x_\beta} + \frac{\partial v_\beta}{\partial x_\alpha}\right ),
\end{equation}
where the incompressible condition, $E_{\alpha\alpha}$=0, is to be satisfied. We can now define a rotation matrix $\Theta$ with which the strain rate tensor $E_{\alpha\beta}$ is transformed to
\begin{equation}\label{eq7}
E'=\Theta E \Theta ^{\rm T} =\left(
\begin{array}{ccc}
0&E'_{xy}&E'_{xz}\\
E'_{yx}&0&E'_{yz}\\
E'_{zx}&E'_{zy}&0
\end{array}
\right),
\end{equation}
where the diagonal components all vanish. 
This transformation makes performing MD simulations much easier with the usual Lees-Edwards periodic boundary condition for simple shear flows under the assumption that each off-diagonal the component of the local stress tensors depends only on the corresponding component of the local strain rate tensors, respectively.
The off-diagonal stress tensor $T'_{\alpha\beta}$ is computed according to $E'_{\alpha\beta}$ and then pass to CFD after transforming back to the original coordinates, $T_{\alpha\beta}$.
For one- or two-dimensional flows [$\partial/\partial z $=0 and $v_z$=0], $\Theta$ and $E'$ are expressed as 
\begin{align}
\Theta&=\left(
\begin{array}{cc}
 \cos\theta& \sin\theta\\
-\sin\theta& \cos\theta
\end{array}
\right),
\label{eq8}
\\
E'_{xy}=E'_{yx}&=-E_{xx}\sin 2\theta + E_{xy}
\cos 2\theta,
\label{eq9}
\end{align}
where
\begin{equation}\label{eq10}
\theta=\frac{1}{2}\tan^{-1}
\left(-\frac{E_{xy}}{E_{xx}}\right).
\end{equation}

Non-equilibrium MD simulations for simple shear flows in the rotated Cartesian coordinates are performed in many MD cells according to the local strain rate $E'$'s defined at each lattice node of the CFD.
The number of particles in each MD cell is 256 if not mentioned.
Once a local stress tensor $P'_{\alpha\beta}$ is obtained at the MD level, the local stress at each lattice node $P_{\alpha\beta}$ in the original coordinate system is obtained by combining the pressure $p$ obtained a priori by CFD and a tensor $T'_{\alpha\beta}$ obtained by subtracting the isotropic normal stress components from $P'_{\alpha\beta}$ as
\begin{equation}\label{eq11}
P = \Theta^{T}[-p {\rm I} + T']\Theta
  =-p {\rm I} +\Theta^{T} T' \Theta,
\end{equation}
where I is the unit tensor. For one- or two dimensional flows, we can use $T'_{xx}$=$T'_{yy}$=0 and $T'_{xy}$=$T'_{yx}$=$P'_{xy}$.

In the non-equilibrium MD simulations, we use the Lees-Edwards sheared periodic boundary condition to a cubic MD. The temperature is kept at a constant by using a thermostat.\cite{art:84BC,book:89AT} The stress $P'_{\alpha\beta}$ is averaged in steady states after transient behavior vanished.

\section{Numerical Computation}
We have carried out the hybrid simulations for one- and two-dimensional flows of a simple liquid composed of the Lennard-Jones (LJ) particles interacting via the potential
\begin{equation}\label{eq12}
v^{\rm LJ}(r)=4\varepsilon\left[
\left(\frac{\sigma}{r}\right)^{12}
-\left(\frac{\sigma}{r}\right)^6
\right].
\end{equation}
In the present simulations, the potential is truncated at $r$=$r_{\rm c}$ and sifted to zero at the distance for computational efficiency.
We considered only the cases where the temperature $T$ and the fluid density $\rho$ are uniform and constant over the CFD systems and the external force is neglected, $g_\alpha$=0. The reduced temperature $T^*$=$Tk/\varepsilon$ and reduced density $\rho^*$=$\rho\sigma^3/m$, where $k$ is the Boltzmann constant and $m$ is the mass of a single LJ particle, are fixed at $T^*$=1.0 and $\rho^*$=0.8 in the simulations. Here and after, non-dimensional quantities normalized by the energy and length parameters of the Lennard-Jones potential, $\varepsilon$ and $\sigma$, are denoted by the superscript ``$\ast$''. 

In the following, $\Delta t$ and $\Delta x$ represent the time-step and the mesh size of CFD calculations, and $t_{\rm MD}$ and $l_{\rm MD}$ represent the sampling time and the side length of a MD cell, respectively.
The two parameters $\Delta t/t_{\rm MD}$ and $\Delta x/l_{\rm MD}$ represent the efficiency of our hybrid simulations. We have carried out hybrid simulations with several different values of the parameters and compared the results with those obtained by usual CFDs. In the present simulations we fixed $t^*_{\rm MD}$=0.005 and $l^*_{\rm MD}$=6.84, while $\Delta t$ and $\Delta x$ are changed as listed in Table \ref{t1}.

\subsection{Pressure-driven channel flows}
The Lennard-Jones liquid with $r^*_c$=2.5 is contained in channel composed
of two parallel plates located at $x_1$=$\pm L/2$ and subjected to a
pressure gradient in $y$-direction. 
We performed one- and
two-dimensional simulations for this pressure-driven channel flows. 
The pressure gradient is set as $\Delta p/(\rho U^2/L)$=1.25, where $\Delta
p$ is the pressure difference over a distance $L$, and $U$ is a
characteristic flow velocity. Non-slip boundary condition is applied on
the two plates.

\begin{figure}[t]
\begin{center}
\includegraphics[scale=1]{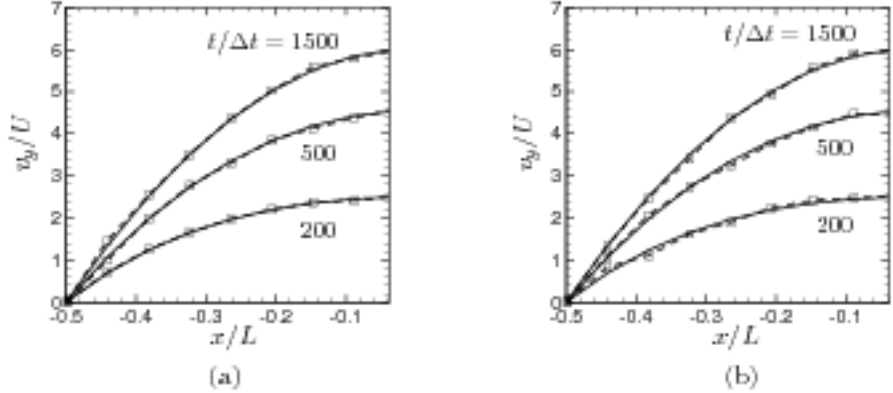}
\end{center}
\caption{The velocity profiles obtained by one dimensional computations for the pressure-driven channel flow. Simulation Parameters are summarized as SP I in Table \ref{t1} for (a) and SP II in Table \ref{t1} for (b).
The solid lines show the results of usual CFD simulation, and dotted lines and square symbols show the results of the present hybrid simulation.}
\label{f1}
\end{figure}

\begin{figure}[th]
\begin{center}
\includegraphics[scale=1]{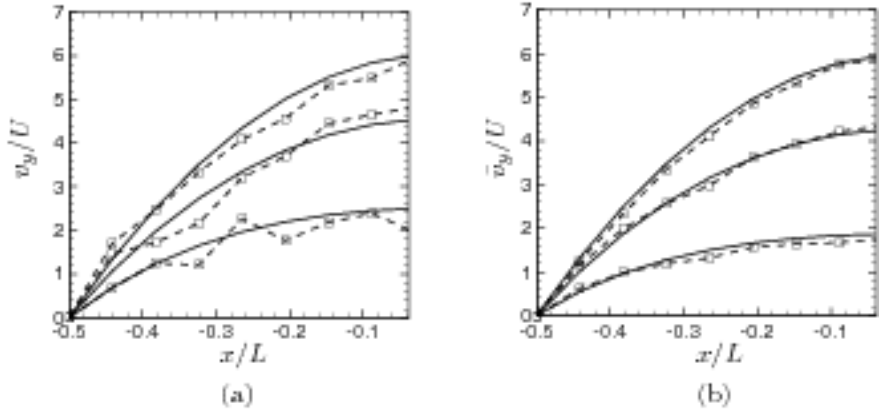}
\end{center}
\caption{The velocity profiles obtained by one dimensional computations for the pressure-driven channel flow. Simulation parameters are summarized as SP III in Table \ref{t1}.
Time evolutions of the velocity profile after an application of pressure gradient in y-direction at $t$=0. Instantaneous profiles at $t/\Delta t=200$, $500$, and $1500$ are shown in (a), while the velocity profiles are time
averaged over $t/\Delta t=[0,300]$, $[300,600]$, and $[1200,1500]$ in (b).
The squares show results of the hybrid simulations, and the dotted lines
present the corresponding CFD results for comparison.
}
\label{f2}
\end{figure}

The results of one-dimensional simulations are shown in Figs.~\ref{f1}
and \ref{f2}. A symmetric condition is used at $x$=0, and the
computational domain [$-L/2$,0] is divided into eight slits. Parameters
used in the simulations are listed as SP I--III in Table \ref{t1}. In
the corresponding CFD simulation, viscosity is set as $\eta^*$=2.0,
which is for the LJ liquid with $r^*_c=2.5$ at $T^*$=1.0 and
$\rho^*$=0.8. The Reynold number defined as $\rho U L/\eta$ is fixed at 40.
It is clearly shown in Fig.~\ref{f1} that the results obtained by the
present hybrid simulations well agree with those of usual CFDs for the
case $\Delta t/t_{\rm MD}<$ 2 and $\Delta x/l_{\rm MD}<$ 2. When the
parameters become large, instantaneous velocity profiles tend to
fluctuate as seen in Fig.~\ref{f2}. It should be noted that the
fluctuations can be removed almost perfectly by taking time
averages. This means that the mean values of the fluctuation is almost
zero, i.e., the fluctuation might be removed also by applying some
filtering, etc. We will discuss on this point later.

\begin{figure}[th]
\begin{center}
\includegraphics[scale=1]{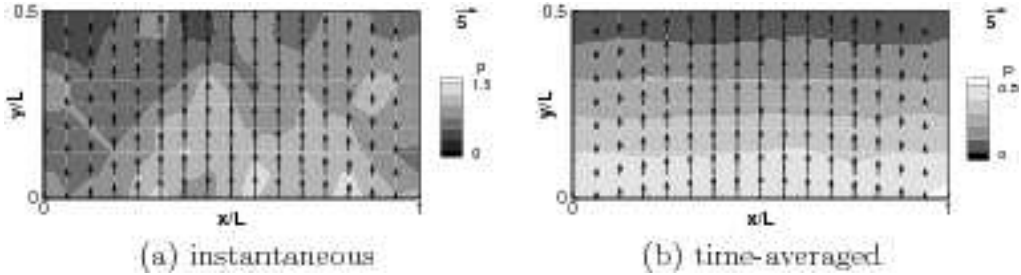}
\end{center}
\caption{The steady-state flow profiles of the pressure-driven channel flow
obtained by a two-dimensional calculation. Simulation parameters are summarized as SP IV in Table \ref{t1}.}
\label{f3}
\end{figure}

The results of two-dimensional simulations are shown in
Fig.~\ref{f3}. The computational domain is now
[-$L/2$,$L/2$]$\times$[0,$L/2$] and divided into 16$\times$8 uniform
lattices. Non-slip boundary condition is used at $x$=$\pm L/2$. At $y$=0
and $L/2$, periodic condition is used for the velocity, and the pressure
is set as $p(x,L/2)$=$p(x,0)-0.5\Delta p$. Here we assumed Stokes flow,
i.e., the second term of the left-hand-side of Equation (\ref{eq2}) is
dropped in the computation. 
It is seen that, at each time step, the velocity fluctuations are much
smaller than the pressure fluctuations. This is clearly due to the
incompressible condition to be imposed to the velocity. The velocity
also fluctuates immediately after solving Eq. (\ref{eq2}), however the
incompressible condition Eq. (\ref{eq1}) tends to adjust it. The
pressure fluctuations can be removed also by taking time averages.

\subsection{Two-dimensional cavity flows}
\begin{figure}[t]
\begin{center}
\includegraphics[scale=1]{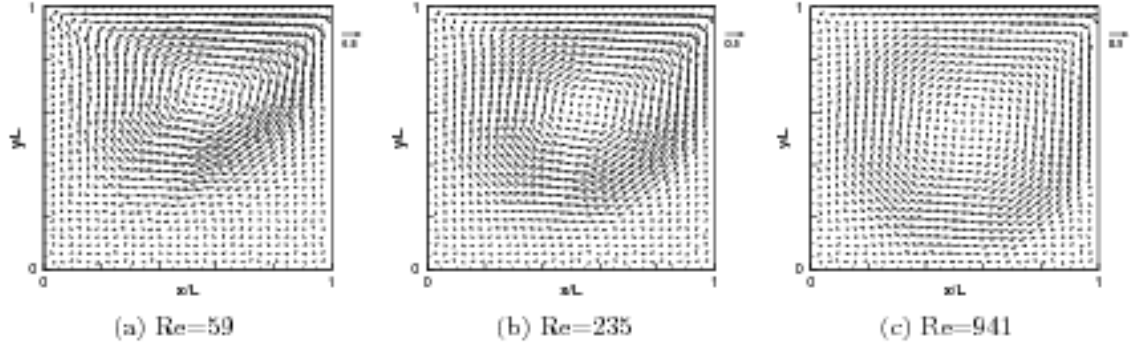}
\end{center}
\caption{The steady-state velocity profile for the cavity flow. Simulation parameters are summarized as SP V in Table \ref{t1} for (a), SP VI for (b), and SP VII for (c). The velocity profiles are time averaged over 
$t/\Delta t$=[950,1000]}
\label{f4}
\end{figure}

\begin{figure}[tbh]
\begin{center}
\includegraphics[scale=1]{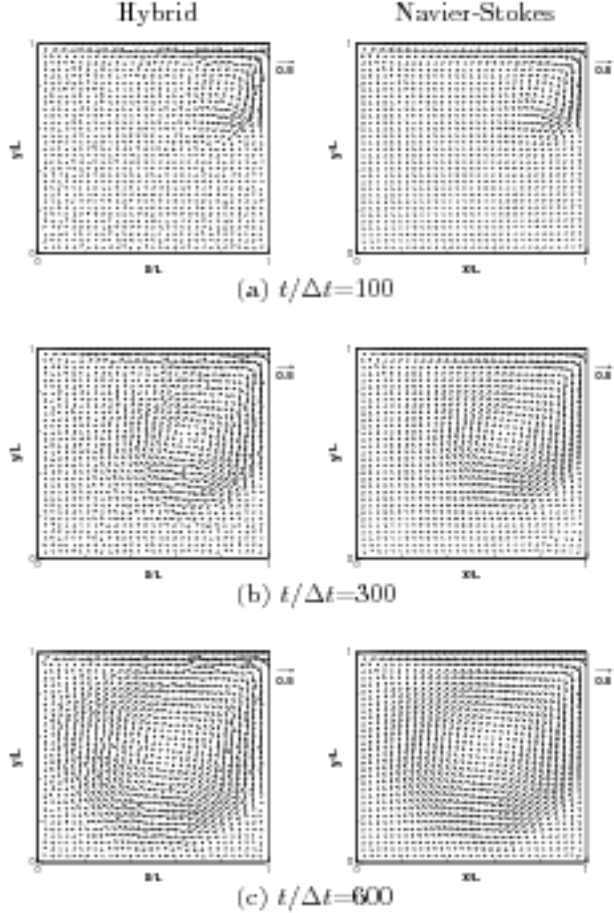}
\end{center}
\caption{Time evolutions of the velocity profile for the cavity flow 
with Re=980. 
The left column shows the Hybrid simulations and the right column shows the corresponding CFD results.
Simulation parameters are summarized as SP VII in Table \ref{t1}.}
\label{f5}
\end{figure}

Lennard-Jones liquid with $r^*_c=2^{1/6}$ is contained in a square box
whose side length is $L$. At $t$=0, the upper wall starts to move from
left to right at a velocity $v_w$=$U$. Non-slip boundary condition is
applied at each wall; $v_x$=$U$ and $v_y$=0 at $y$=$L$, and
$v_x$=$v_y$=0 at other walls. At left- and right- upper corners,
$v_x$=$U$ and $v_y$=0 is applied.
The results of the hybrid simulations are shown in Figs.~\ref{f4} and
\ref{f5}. The computational domain is divided into 32$\times$32 uniform
lattices. Values of parameters used in the present simulations are
listed as SP V-VII in Table \ref{t1}. The Reynolds number is defined as
$\rho U L /\eta$, and the viscosity of the corresponding LJ fluid is
$\eta^*$=1.7.

Fig.~\ref{f4} shows the steady-state velocity profiles time-averaged
over $t/\Delta t$=[950,1000]. It is clear that our hybrid method can
successfully reproduce the characteristic flow properties of cavity
flows with different Reynolds numbers. 
Fig. \ref{f5} shows time evolutions of velocity profiles for the case
of Re=980 after a sudden application of upper-wall sliding at $t=0$. 
Here, the results
obtained by hybrid simulations are compared with those of usual CFD
simulations. It is seen commonly that a small vortex first appears at
the upper-right corner is moving gradually toward the center of the box
with increasing the size of the vortex as time passes. 
The agreements between hybrid simulations and CFDs are very well.
Our hybrid method is confirmed to reproduce successfully the
time-evolution while large fluctuations are seen in the instantaneous
velocity profiles. 
\begin{table}[t]
\begin{center}
\begin{tabular}{c cc cc cc cc cc cc}
\hline\hline
\multicolumn{13}{c}{1d channel flow}\\
\hline
    && $l^*_{\rm MD}$
    && $t^*_{\rm MD}$
    && $L^*$
    && $U^*$
    && $\Delta x/l_{\rm MD}$
    && $\Delta t/t_{\rm MD}$\\
    \hline
SP I   && 6.84 && 1.87 && 116.3 && 0.86 && 1.0  && 1.0  \\
SP II  && 6.84 && 3.74 && 232.6 && 0.43 && 2.0  && 2.0  \\
SP III && 6.84 && 7.49 && 465.1 && 0.22 && 4.0  && 4.0  \\ 

\multicolumn{13}{c}{2d channel flow}\\
\hline
    && $l^*_{\rm MD}$
    && $t^*_{\rm MD}$
    && $L^*$
    && $U^*$
    && $\Delta x/l_{\rm MD}$
    && $\Delta t/t_{\rm MD}$\\
    \hline
SP IV  && 6.84 && 3.74 && 232.6 && 0.43 && 2.0  && 2.0  \\

\multicolumn{13}{c}{2d cavity flow}\\
\hline
    && $l^*_{\rm MD}$
    && $t^*_{\rm MD}$
    && $L^*$
    && $U^*$
    && $\Delta x/l_{\rm MD}$
    && $\Delta t/t_{\rm MD}$\\
    \hline
SP V    && 6.84 && 4.68 && 218.9 && 0.46 && 1.0  && 1.0 \\
SP VI   && 6.84 && 9.36 && 437.8 && 0.91 && 2.0  && 2.0 \\
SP VII  && 6.84 && 3.11 && 875.5 && 1.83 && 4.0  && 4.0 \\
\hline\hline
\end{tabular}
\caption{Simulation parameters.}
\label{t1}
\end{center}
\end{table}

\section{Discussion}
As mentioned above, the ratios $\Delta t/t_{\rm MD}$ and $\Delta
x/l_{\rm MD}$ measure the efficiency of our hybrid simulations. Larger
the ratios, simulations are more efficient, however, the statistical
fluctuations also become large. For example, in a case of $\Delta
t/t_{\rm MD}$=$\Delta x/l_{\rm MD}$=4, computational efficiency is,
roughly speaking, $4^D\times 4$ times more efficient than a full MD
simulation of a $D$-dimensional cubic system. As we have already seen in
one- and two-dimension cases, numerical results of our hybrid
simulations show good agreements with those of CFD simulations as far as
$\Delta t/t_{\rm MD}$ and $\Delta x/l_{\rm MD}$ remain small, 
say $\Delta t/t_{\rm MD} < 2$ and $\Delta x/l_{\rm MD}< 2$. 
In fact, the normalized standard deviation,
$\int_Ldx\int_Tdt(v-v_{\rm NS})^2/TL$, of the velocity profiles of
hybrid method, $v$, and those of CFDs, $v_{\rm NS}$, are less than 0.02
and 0.07 for the cases of Fig.~\ref{f1} (a) and for Fig.~\ref{f1} (b).
As the ratios increase, solutions of our hybrid model start to fluctuate
around the corresponding CFD results. The deviation becomes about 0.6 in
the case of Fig.~\ref{f2} (a). 
It is worth mentioning that the instantaneous velocity fluctuations are
notable at each time step, however, they can be removed almost perfectly
by taking time averages.
In the following part, we will discuss the nature of the fluctuations 
more in detail to examine possibilities of effectively controlling them 
in our future simulations where correct thermal fluctuations will be included.

To handle the statistical noise explicitly, we rewrite Eq. (\ref{eq11}) as
\begin{equation}\label{eq13}
P=-pI+\Theta^T(T_*'+R')\Theta,
\end{equation}
where the off-diagonal stress tensor $T'$, which is to be determined by
MD sampling, is decomposed into the non-fluctuating stress $T'_*$ and
the fluctuating random stress $R'$ due to the thermal noise. 
The magnitude of each component of the random stress included in MD sampling 
$\langle R_{{\rm MD}pq}'^2\rangle$, where $p$ and $q$ represent
the index in Cartesian coordinates  and do not follow the summation convention,
should depend both on the size of the MD cell $l_{\rm MD}$ and
the length of time $t_{\rm MD}$ over which average is taken at the
MD level; 
$\langle R_{{\rm MD}pq}'^2\rangle$=$\langle \bar R_{pq}(l_{\rm MD},t_{\rm MD})^2\rangle$,
where $\bar R(l,t)$ represents the random stress tensor averaged in a cubic with a side length $l$ and over a time duration $t$. 

At the CFD level which is discretized with a mesh size $\Delta
x$ and a time-step $\Delta t$, the physically correct magnitude should
be 
$\langle R_{{\rm CFD}{pq}}'^2\rangle$
=$\langle \bar R_{pq}(\Delta x,\Delta t)^2\rangle$.
If the central limit theorem,
$\langle\bar R_{pq}(l,t)^2\rangle$$\propto 1/l^D t$ is assumed, 
the following simple formula can be used. 
\begin{equation}\label{eq14}
\langle R_{{\rm MD}{pq}}'^2\rangle=
\left( \frac{\Delta x}{l_{\rm MD}} \right)^D
\left(\frac{\Delta t}{t_{\rm MD}} \right)
\langle R_{{\rm CFD}{pq}}'^2\rangle.
\end{equation}
This finally leads to the following very useful expression for the
correctly fluctuating stress tensor $P$,
\begin{equation}\label{eq15}
P = -pI+\Theta^T\left[
T'_*+ 
\sqrt{
\left(\frac{l_{\rm MD}}{\Delta x}\right)^D
\left(\frac{t_{\rm MD}}{\Delta t}\right)
}
R'_{\rm MD}
\right] \Theta
\end{equation}
to be used in CFD instead of Eq.~(\ref{eq11}).
This equation indicates that if we can re-weight randomly fluctuating
part ${R}'$ while the non-fluctuating part $T'_*$ being untouched, 
hydrodynamic simulations including correct thermal fluctuations can be
done for complex fluids within the present framework.

We note that the important key toward the development of 
fluctuating hybrid simulation is the separation of $T'_*$ and ${R}'$.
We thus carried out spectral analysis for the fluctuations in the total 
stress tensor computed directly from MD simulations $T'=T'_* + {R}'$. 
The discrete Fourier transformation of $T'_{xy}$ is defined as
\begin{equation}
\varPi'_{xy}\{\bm k\}=\frac{1}{4M^2}\sum_{n_x=0}^{2M-1} \sum_{n_y=0}^{2M-1} \hat T'_{xy}\{\bm x\}\exp(-i{\bm k}\cdot{\bm x}),
\end{equation}
where ${\bm x}=(n_x\Delta x, n_y\Delta x)$ is the position of each lattice
node $(n_x,n_y)$, ${\bm k}=(2\pi m_x/L, 2\pi m_y/L)$ is
the wave vector, $n_x,n_y,m_z,m_y$ are integers,
M is the lattice number in each $x$- and $y$-axis,
and $\hat T'_{xy}\{ {\bm x} \}$ is defined as
$\hat T'_{xy}\{ {\bm x} \}$=$T'_{xy}(x+\Delta x/2,y+\Delta x/2)$
for $0\le x, y \le L$, 
$\hat T'_{xy}\{ {\bm x} \}$=$T'_{xy}\{2L-x,y\}$ for $L<x\le 2L$, 
and
$\hat T'_{xy}\{ {\bm x} \}$=$T'_{xy}\{x,2L-y\}$ for $L<y\le 2L$.

\begin{figure}[tb]
\begin{center}
\includegraphics[scale=1]{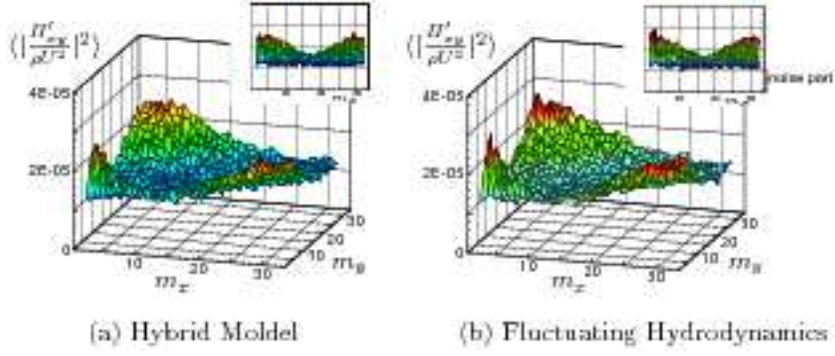}
\end{center}
\caption{
The fluctuations of $T'_{xy}$ for the case of cavity flow with Re=59. 
The power spectra $\langle |\varPi'_{xy}\{\bm k\}|^2\rangle$ for the 
present multi-scale model with $\Delta x/l_{\rm MD}=\Delta t/t_{\rm MD}=1$ 
is shown in (a) and the corresponding result from the fluctuating 
hydrodynamics is shown in (b) for a comparison.
$\varPi'_{xy}$ represents the discrete Fourier transform of $T'_{xy}$.
$m_\alpha$ is defined as $m_\alpha$=$(L/2\pi)k_\alpha$, where ${\bm k}$ is the wave vector.
The insets on each figure shows the $\langle|\varPi'_{xy}|^2\rangle$-$m_x$ plane.
}
\label{f6}
\end{figure}

The power spectra $\langle |\varPi'_{xy}\{\bm k\}|^2\rangle$ calculated 
from our hybrid simulations of driven cavity flows  
are plotted in Fig. \ref{f6} (a) 
for the case of $\Delta t/t_{\rm MD}=\Delta x/l_{\rm MD}=1$.
This corresponds to the case of Fig. \ref{f4} (a).
The angle bracket $\langle\cdots\rangle$ means the time average 
taken in the steady state at CFD level.
One can see that the overall structure is rather simple. 
There exists a relatively large peak around ${\bm k}=0$
and rather flat distributions throughout the ${\bm k}$ plane.
The former corresponds to the contributions from the non-fluctuating 
part $T'_*$ and the later corresponds to the contributions from the random
stress $R'$.
The same quantity obtained by conventional fluctuating hydrodynamics 
using a constant Newtonian viscosity and the random stress whose
intensity is determined by the fluctuation-dissipation
theorem\cite{book:59LL} is shown in Fig. \ref{f6} (b) for a comparison.\cite{note1}
Those two plots are surprisingly similar to each other including the
fluctuation part. 
This means that our hybrid simulation generates 
fluctuations quite consistent with the fluctuating hydrodynamics with 
fluctuation-dissipation theorem
in the case of $\Delta x/l_{\rm MD}$=$\Delta t/t_{\rm MD}$=1.

\begin{figure}[tb]
\begin{center}
\includegraphics[scale=1]{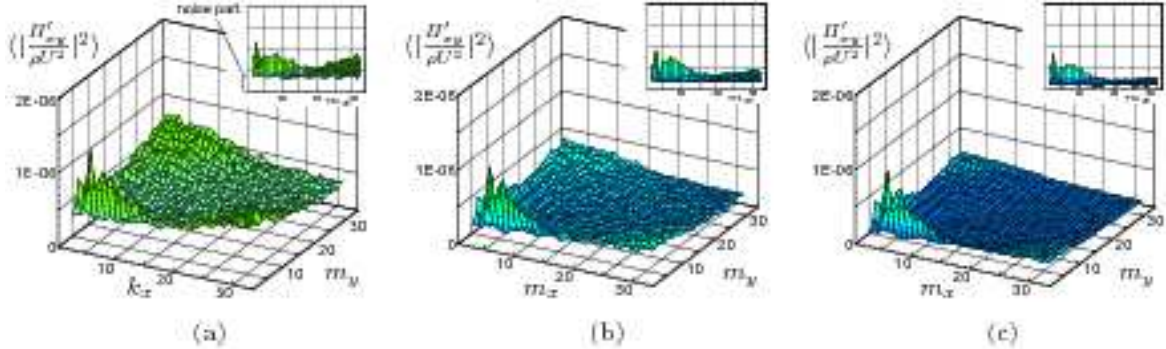}
\end{center}
\caption{
The fluctuations of $T'_{xy}$ for the case of cavity flow with Re=235. 
The power spectra $\langle |\varPi'_{xy}\{\bm k\}|^2\rangle$ is plotted in (a) 
for the case of Fig. \ref{f4} (b).
Only the number of particles are doubled in (b), while other parameters 
are unchanged from (a).
In (c), both the number of particles and the sampling time of $T'_{xy}$ 
are doubled. 
$\varPi'_{xy}$ represents the discrete Fourier transform of $T'_{xy}$. 
$m_\alpha$ is defined as $m_\alpha$=$(L/2\pi)k_\alpha$, where ${\bm k}$ is the wave vector.
The insets on each figure shows the $\langle|\varPi'_{xy}|^2\rangle$-$m_x$ plane.
}
\label{f7}
\end{figure}

Next, one see how the fluctuations depend on the ratios 
$\Delta x/l_{\rm MD}$ and $\Delta t/t_{\rm MD}$ in Fig. \ref{f7}.
Here, comparing to the reference case (a) 
[$\Delta x/l_{\rm MD}$=$\Delta t/t_{\rm MD}$=2],
the number of particles used in MD simulations are doubled in the case of (b)
[$\Delta x/l_{\rm MD}$=1.59, $\Delta t/t_{\rm MD}$=2],
and both the number of particles and the sampling duration 
to take time average are doubled in the case of (c)
[$\Delta x/l_{\rm MD}$=1.59, $\Delta t/t_{\rm MD}$=1].
It is seen that the noise intensity decreases with decreasing ratios
$\Delta x/l_{\rm MD}$ and $\Delta t/t_{\rm MD}$
in a consistent way to the central limiting theorem Eq. (\ref{eq14})
{\it i.e.}, the noise intensity in (b) is about a half of that in (a),
and the intensity in (c) is about one fourth of that in (a).
%
%

Finally, we mention other recently proposed methods based on a similar idea.
In the reference \onlinecite{art:05RE},
a hybrid method is proposed for bulk and boundary problems. 
Several problems for one- or two-dimensional flows of simple
Lennard-Jones and dumb-bell liquids are considered.
We note that the present multi-scale hybrid method is different from 
the methods proposed in those references particularly on the
constructions of the stress tensor. 
In our method, a rotation matrix which effectively transforms 
the tensors in the Cartesian coordinates used in CFD and MD simulations.
We also replace the isotropic part of the stress tensor calculated
by MD simulations with the pressure imposed by the incompressible
condition in CFD. More specifically, only the pure shear stress is
passed from MD to CFD for numerical efficiency and consistency.

\section{Summary}
We proposed a multi-scale method for hybrid simulations of MD and CFD. 
Our method is based on direct computations of the local stress by
performing non-equilibrium MD simulations according to the local flow
field at all lattice nodes of CFD. The validity of the method is tested
by comparing the numerical results obtained by our method and usual CFD.
We found that the results obtained by our hybrid method agree well with
those of usual CFDs with the Newtonian constitutive relation when the
mesh size and the time-step of CFD are not too large comparing to the
cell size and sampling time of MD simulations. 
When the ratios $\Delta t/t_{\rm MD}$ and $\Delta x/l_{\rm MD}$ become large, there appear large fluctuations in flow field of our hybrid simulations.
It was, however, clarified by the spectral analysis that the stress
tensor $T'$ computed by MD simulations has a very simple structure. It
is composed of the non-fluctuating component $T'_*$ and the random
component ${R'}$ which seem to obey simple central limiting theorem
according to the system size and the duration of the MD sampling.
We confirmed that the power spectrum of the non-fluctuating component $T_*'$ in MD sampling agrees well with that computed in usual CFD without fluctuation.
The power spectrum of ${R'}$ also showed a good agreement with numerical results of the fluctuating hydrodynamics which obeys the fluctuation-dissipation theorem.

\section*{Acknowledgment}
The authors would like to express their gratitude to
Professor Weinan E for for useful discussions.



\begin{thebibliography}{99}
\bibitem{note1}
In the present simulations of fluctuating hydrodynamics, the random noises are included only in the off-diagonal components of the stress tensor $T'$. That is, we put $T'_{xx}=T'_{yy}=0$ and 
$T'_{xy}$=$T'_{yx}$=$2\eta E'_{xy}$+$R'_{{\rm CFD}{xy}}$,
where the intensity of the random noise is defined by the Fluctuation-dissipation theorem as $\langle R_{{\rm CFD}{xy}}'^2\rangle = 2\eta kT/(\Delta x^3\Delta t)$.
\end{thebibliography}

\end{document}